\documentclass[pre,eps,aps,reprint,twocolumn,superscriptaddress]{revtex4-2}
\usepackage{graphicx}
\usepackage{amsmath}
\usepackage{amssymb}
\usepackage{color}
\usepackage[latin2]{inputenc}
\def\slaninafigdir{.}
\begin{document}
\title{%
Short-range and long-range correlations in driven dense colloidal mixtures in narrow pores
}
\author{%
Franti\v{s}ek Slanina%
}%
\affiliation{%
Institute of Physics,
 Czech Academy of Sciences,
 Na~Slovance~2, CZ-18221~Praha,
Czech Republic%
}%
\email{
slanina@fzu.cz
}%
\author{%
Miroslav Kotrla%
}%
\affiliation{%
Institute of Physics,
 Czech Academy of Sciences,
 Na~Slovance~2, CZ-18221~Praha,
Czech Republic%
}%
\author{%
Karel Neto\v{c}n\'y%
}%
\affiliation{%
Institute of Physics,
 Czech Academy of Sciences,
 Na~Slovance~2, CZ-18221~Praha,
Czech Republic%
}%
\begin{abstract}
The system of driven dense colloid mixture in a tube with diameter
comparable with particle size is modeled by a generalization of
asymmetric simple exclusion (ASEP)
model. The generalization goes in two directions: relaxing the
exclusion constraint by allowing several (but few) particles on a
site, and by considering two species of particles, which differ by
size and transport coefficients. We calculate the nearest-neighbor
correlations using a variant of Kirkwood approximation and show by
comparison with numerical simulations that the approximation provides
quite accurate results. However, for long-range correlations, we show
that the Kirkwood approximation is useless, as it predicts exponential
decay of the density-density correlation function with distance, while
simulation data indicate that the decay is 
algebraic. For one-component system, we show that the decay is
governed by a power law with universal exponent close to $2$. In two-component system,
the correlation function behaves in more complicated manner; its sign
oscillates but the envelope decays again very slowly and the decay is
compatible with power-law with exponent somewhat lower than $2$.
Therefore, our generalization of ASEP belongs to different
universality class than the ensemble of generalized ASEP models which
are mappable to zero-range processes.
\end{abstract}
%
%
\maketitle%
\section{Introduction}

Dynamics of colloidal suspensions becomes complex when the particles
move in geometrically constrained volume at high concentrations. A
typical situation of this kind is pushing the suspension through a
membrane pierced by a number of pores, aperture of which is comparable
with particle size \cite{mat_mul_03}.
In the extreme case, just single particle can pass
through the pore simultaneously, which results in a single-file like
dynamics \cite{wei_bec_lei_00}. In slightly less strict confinement,
the aperture can
accommodate several particles, but still a small number of them.

A simple model well suited for describing such situations is the asymmetric
simple exclusion model (ASEP) \cite{derrida_98} and its
generalizations. Introduced
originally in the context of ribosome movement along the RNA chain
\cite{mcd_gib_pip_68}, ASEP became widely popular for its unique
features. The most important, from practical point of view, is, that
it is exactly solvable, both with open boundary conditions
and with
periodic boundary conditions
\cite{do_do_mu_92,de_ev_pa_93,de_ja_le_spe_93a,schutz_93a,sandow_94,sch_zia_95,schutz_97,schutz_01,bly_eva_07}. From a fundamental point of view, the
stunning property of ASEP is the absence of correlations in stationary
state. Therefore, for example, a single-line calculation of mean-field
type gives
exact current-density diagram.
Of course, as soon as we depart from stationarity, the full complexity
of ASEP comes into play and reveals non-trivial exponents, for example
when mapped on surface growth models \cite{gwa_spo_92b}.
But still, when we remain in stationary state, we can pose a natural
question, how robust is the absence of correlations when we modify the
rules of ASEP in this or that manner. It is tempting to expect that,
when the correlations appear, they are short ranged or perhaps reducible to
nearest-neighbor correlations.

One of the ways to generalizations of ASEP is modification of hopping
probabilities for a particle, depending on the configuration of
particles behind and/or in front of it. For example, hopping rate may
be enhanced by a factor, if there is a particle just behind the
hopping particle. This is the so-called facilitated ASEP model (the
name comes from the situation where the factor is greater than one,
although generally the factor can be arbitrary positive number)
\cite{gab_kra_red_10,bas_moh_09}. This model is also exactly soluble
and the pair
correlation function can be obtained explicitly. Its sign oscillates
and absolute value decays exponentially with distance. Moreover, it
can be shown that the two-site cluster mean field approximation gives exact
result \cite{gab_kra_red_10,bas_moh_09,pin_gov_13}.
This is the fortunate case when all information on
correlations is obtained in the nearest-neighbor correlation
function.

Similarly, the
hopping rate can depend on the number of free sites either in front or
behind the particle \cite{bas_moh_10a,pri_ayy_jai_14}, or particle
possess an internal degree of
freedom \cite{bas_moh_10}.
(In fact, the facilitated ASEP is a special case
of this family of models.) Here again, exact solution is available,
using mapping on zero-range processes \cite{eva_han_05} and the
correlation function decays exponentially with distance. Analogous
result  was
found in an exactly soluble variant in which particles interact
by a particular type of repulsive interactions \cite{krapivsky_13}.
It is interesting to note that the system of hard rods on a line, the
so-called Tonks gas, was solved long time ago
\cite{tonks_36,sal_zwa_kir_53}. In the context of generalized ASEP
models it corresponds to the model of driven k-mers, which was also
solved exactly \cite{gup_bar_bas_moh_11} and the oscillating character
of pair correlation function closely resembles that of the Tonks gas.

However, for example in \cite{mid_kol_gup_18} they study a generalizations of
ASEP
which cannot be mapped on
a zero-range process and show that the correlations are not reducible
to nearest-neighbor ones. Hence the two-site cluster mean-field
approximation is no more exact, although it seems quite accurate.

The exact results of \cite{gab_kra_red_10,bas_moh_09,pin_gov_13} were then
used and together with simulations of wider class of related models
served to the conjecture that exponential decay of correlation is a
universal feature \cite{hao_jia_hu_jia_wan_16,hao_che_sun_liu_wu_16}.

Let us also mention the relation to traffic-flow models, which differ
from ASEP and its variants by parallel-update dynamics
\cite{na_sch_92}. Here also, the correlation function shows exponential
decay with oscillating sign
\cite{che_ker_sch_98,neu_lee_sch_99,cho_san_sch_00} and
two-point cluster mean-field approximation provides exact result
\cite{sch_sch_93}.

The generalization of ASEP which will be mostly relevant to our work
consists in relaxing the exclusion principle. Instead of allowing just
single particle on a site, we allow more, but at most $k$ particles on
site. Exclusion model with this constraint was introduced in
\cite{kip_lan_oll_94} in a symmetric variant (i.e. hopping without
bias).
It was thoroughly studied in \cite{ari_kra_mal_17}, where also nearest-neighbor
correlations were calculated by numerical simulations. This family of
generalized exclusion process was also investigated from the
point of view of hydrodynamic limit  \cite{seppalainen_99} and
effective transport
coefficients
\cite{bec_nel_cle_par_bro_13,bec_nel_cle_par_bro_14,ari_kra_mal_14,ari_kra_mal_17}.

In our work, we shall study the variant of ASEP generalized in two
ways. First, we allow that one site accommodates more than one
particle, as in \cite{kip_lan_oll_94}, second, we shall work with more
particle types (specifically, two types) which differ by size and
hopping parameters. These generalizations were introduced and studied
in our previous work \cite{hum_kot_net_sla_20,hum_kot_sla_21}, where
the focus was on ratchet
effect. In this work, we concentrate instead on correlation properties
and we aim at showing that our generalization of ASEP differs in this
respect from the generalized ASEP versions mentioned above.

In calculation of the short-range correlations we shall rely on the
Kirkwood approximation. Originally, it was developed for calculation
of properties of liquid mixtures \cite{kirkwood_35} and dense fluids
in general \cite{kir_lew_ald_52}.
 {
Subsequently, it was used in various
contexts, notably, for our  purposes, in one-dimensional stochastic
many-particle systems, see
e. g. \cite{sch_sch_nag_ito_95,mob_red_03,sla_szn_prz_08}.
The information-theory analysis hinted \cite{matsuda_00} that from
certain point of view it is the optimal decoupling approximation for
three-site correlation function.
}

The results on models investigated in
\cite{gab_kra_red_10,bas_moh_09,bas_moh_10,bas_moh_10a,pri_ayy_jai_14}
suggest that in situations where Kirkwood approximation becomes exact,
the pair correlation function decays exponentially with distance. This
can be intuitively understood when we imagine the long-range
correlation function as composed of chained Kirkwood decouplings into
a product of nearest neighbor functions. If Kirkwood approximation is
actually exact, also this chained decoupling is exact and its
multiplicative structure implies exponential decay of
correlations. More precisely, correlation function is expressed in
terms of a matrix to a power (equal to distance) and therefore it is a
sum of a few exponentials. However, in our work we shall try to show
that  in our generalization of ASEP this scenario is inapplicable and
the correlations decay algebraically, rather than exponentially. In
fact, algebraic decay if correlation was also observed in
a variant of ASEP studied in \cite{pri_ayy_jai_14}, but only for a
single critical value of a parameter of the model.

\section{Local mixing approximation and generalized ASEP model}

Let us consider the system of spherical colloid particles interacting
by steric repulsion, enclosed in a straight tube  of diameter
$d$. There are $M$ species of particles, distinguished by their
diameter $d_\alpha$, $\alpha=1,2,\ldots,M$. The motion of each
particle taken individually is due to Brownian motion with a bias. For
the coordinate $x_{\alpha i}$ of the center of $i$-th particle of type
$\alpha$ we have standard equation
\begin{equation}
\mathrm{d} x_{\alpha i}(t)=f_\alpha\mathrm{d} t
+\mathrm{d} W_{\alpha i}(t)
\label{eq:stochcontinuous}
\end{equation}
where $W_{\alpha i}(t)$ is ensemble of independent Wiener
processes,
$(\mathrm{d}W_{\alpha i})^2=2D_\alpha\mathrm{d} t$.  {The diffusion
coefficient $D_\alpha$ and the drift $f_\alpha$ depend only on the
particle type $\alpha$. }
(Here and in the following we write the model equations in one
spatial dimension, but we have in mind general case of arbitrary
spatial dimension.)  The steric constraints
\begin{equation}
|x_{\alpha i}-x_{\beta j}|> \frac{1}{2}(d_{\alpha }+d_{\beta })
\label{eq:stericcontinuous}
\end{equation}
valid at all times make the stochastic dynamics non-trivial.
 {To help to develop an intuition, a common physical system
can be kept in mind, namely blood in capillary veins or in
laboratory microfluidic chambers. Blood can be
considered as dense mixture of a few species of colloidal particles
which differ in size and shape. Purely physical methods of
separation of blood particles from the mixture are developed and
used in practice, see e. g. \cite{dic_iri_tom_ton_07}}

To tackle the problem we
first replace the spatially continuous dynamics by discrete one. To
this end, we partition the space into disjoint cells and neglect the
dynamics of the particles within the cells. The only thing which
remains of the stochastic process (\ref{eq:stochcontinuous}) for
independent particle is random
hopping between cells
\begin{equation}
x_{\alpha i}(t) - x_{\alpha i}(0) = S_+(t)-S_-(t)
\label{eq:stochdiscrete}
\end{equation}
where $S_+(t)$ and $S_-(t)$ are Poisson processes with rates
$a_\alpha$ and $b_\alpha$, respectively, which are specific to each
particle type. They are related to the
properties of the process  (\ref{eq:stochcontinuous}) as
$a_\alpha-b_\alpha=f_\alpha$, $a_\alpha+b_\alpha=2D_\alpha$.
For simplicity, we fix the unit length as the cell size.

Moreover, the steric constraint
(\ref{eq:stericcontinuous}) is taken into account by the requirement
that only certain configurations of particles can enter into the
cell. More specifically, we fix a cell capacity $k$ and weight factors
$c_\alpha$ which are related to particle diameters $d_\alpha$, so that
the numbers $n_\alpha$ of particles of type $\alpha$ within one cell
must satisfy
\begin{equation}
\sum_{\alpha=1}^M c_\alpha n_\alpha \leq k \;.
\label{eq:stericdiscrete}
\end{equation}
Trajectories produced by the process (\ref{eq:stochdiscrete}) but
violating the constraint (\ref{eq:stericdiscrete}) (which must be
satisfied in every cell at all times) are forbidden.

Such an approximation effectively assumes that the dynamics within
cell is fast, so that the particles are mixed on the level of cells
and for description of the global behavior of the mixture it is
sufficient to consider inter-cell hopping constrained by the condition
(\ref{eq:stericdiscrete}). We can call such approach the local mixing
approximation. There is a significant level of arbitrariness in the
choice of the cell size. Certainly, larger cell means larger $k$ and
to certain extent also change in the weights $c_\alpha$. In order to
establish with certainty how many particles can enter a cell of given
size and shape, one needs to solve a very complex
constraint-satisfaction problem of sphere packing
\cite{con_slo_99}. Moreover, as particle configurations
are result of random process, regularly ordered configurations, which
are the most space-saving, occur with prohibitively small
probabilities. Therefore, we are left with random close
packings\cite{ber_mas_60,tor_sti_10}, or
hard-sphere glasses \cite{par_zam_10,cha_kur_par_urb_zam_14},
which are still far from being completely known (see e.g.
\cite{liu_nag_10,liu_nag_saa_wya_11,cha_cor_par_zam_12,wyart_12,nes_cat_20,ris_cor_par_21}).

Here we skip all these hard questions and solve the dynamics for
several cases fixed by the choice of the parameters $k$, $c_\alpha$,
and particle-hopping rates. The cell size enters just through the
parameter $k$ and the uncertainty about the proper choice of the cell
size will be counterweighted by keeping just those conclusions, which
depend weekly, or in well-controlled manner, on the choice of $k$.

 {
In practice, the appropriate choice of $k$ must be based on comparison
of the results  of the continuous process  (\ref{eq:stochcontinuous})
supplemented by constraints (\ref{eq:stericcontinuous}),
with the generalized ASEP process (\ref{eq:stochdiscrete}) with
constraints (\ref{eq:stericdiscrete}). We shall show an example of
such comparison later.}

The simplest case $k=1$ with only one type of particles, $M=1$ and
put on a linear chain is the well-known ASEP model, which is exactly
soluble. However, as soon as $k>1$, the system is no more integrable
and we must resort to approximations and Monte Carlo simulations. In our
previous works, we
investigated various features of these generalized ASEP models, mainly
using mean-field approximation
\cite{hum_kot_net_sla_20,hum_kot_sla_21}. We found that in systems
composed of one type of particles only, the mean-field approximation
provides fairly good results for particle current.  At the same time, we found
that in the case of mixtures, mean-field is significantly less
reliable when compared with Monte Carlo results. Therefore,
correlation effects are significant. Fortunately enough, for
computation of particle current it is necessary to know just
nearest-neighbor correlations. There is a well-established
approximation scheme called Kirkwood approximation
\cite{kirkwood_35,kir_lew_ald_52,matsuda_00,sch_sch_nag_ito_95,mob_red_03,sla_szn_prz_08}
for
calculating these correlations. In the next section we shall expose
how we adapt the Kirkwood approximation for our family of generalized
ASEP models.

\section{Kirkwood approximation in generalized ASEP models}

To take a simplest non-trivial example,
we shall deal with system composed of two types of particles
($M=2$) which we shall call small and big ones. Their size will be
characterized by factors $c_1=1$ and $c_2=2$. The cell capacity will
be $k$ so that the number of small $n_s$ and number of big $n_b$
particles in cell indexed by $x$ must satisfy
\begin{equation}
n_s(x)+2n_b(x)\le k\;\;\quad\forall x\;.
\label{eq:stericbigsmall}
\end{equation}
The positions of the particles will evolve according to process
(\ref{eq:stochdiscrete}) with constraint (\ref{eq:stericbigsmall})
satisfied at all times. To simplify a bit the notation, the hopping
rates $a_\alpha,b_\alpha$ will be denoted $a$ and $b$ for small
particles and $A$ and $B$ for big particles.
The cells form a one-dimensional regular lattice of length $L$. This
corresponds to particles of the original model
(\ref{eq:stochcontinuous}), (\ref{eq:stericcontinuous}),
constrained to a straight tube of constant diameter. We assume periodic boundary conditions.
There are in
total $N_s$ small and $N_b$ big particles, so the average density is
$\rho_s=N_s/L$ for small particles and $\rho_b=N_b/L$ for big particles.

We shall investigate the local one-, two-, and three-site local probabilities
\begin{equation}
\begin{split}
P^{(1)}_{nm}=\frac{1}{L}\sum_{x=1}^L\mathrm{Prob}\{n_s(x)=n,n_b(x)=m\}\\
P^{(2)}_{nmn'm'}=\frac{1}{L}\sum_{x=1}^L
\mathrm{Prob}\{n_s(x)=n,n_b(x)=m,\\
n_s(x+1)=n',n_b(x+1)=m'\}\\
P^{(3)}_{nmn'm'n''m''}=\frac{1}{L}\sum_{x=1}^L
\mathrm{Prob}\{n_s(x)=n,n_b(x)=m,\\
n_s(x+1)=n',n_b(x+1)=m',\\
n_s(x+2)=n'',n_b(x+2)=m''\}\;.
\end{split}
\end{equation}
The one-site and two-site probabilities evolve in time according to the following
equations
\begin{widetext}
\begin{equation}
\begin{split}
\frac{\mathrm{d}}{\mathrm{d}t}P^{(1)}_{nm}=
\sum_{n'm'}\Bigg[
&an'P^{(2)}_{n'm'n-1m}+bn'P^{(2)}_{n-1mn'm'}+\\
&a(n+1)P^{(2)}_{n+1mn'm'}\widetilde{\delta}(n',m')+
b(n+1)P^{(2)}_{n'm'n+1m}\widetilde{\delta}(n',m')-\\
&anP^{(2)}_{nmn'm'}\widetilde{\delta}(n',m')-
bnP^{(2)}_{n'm'nm}\widetilde{\delta}(n',m')-
an'P^{(2)}_{n'm'nm}\widetilde{\delta}(n,m)-
bn'P^{(2)}_{nmn'm'}\widetilde{\delta}(n,m)+\\
&Am'P^{(2)}_{n'm'nm-1}+BmP^{(2)}_{nm-1n'm'}+\\
&A(m+1)P^{(2)}_{nm+1n'm'}\widetilde{\Delta}(n',m')+
B(m+1)P^{(2)}_{n'm'nm+1}\widetilde{\Delta}(n',m')-\\
&AmP^{(2)}_{nmn'm'}\widetilde{\Delta}(n',m')-
BmP^{(2)}_{n'm'nm}\widetilde{\Delta}(n',m')-
Am'P^{(2)}_{n'm'nm}\widetilde{\Delta}(n,m)-
Bm'P^{(2)}_{nmn'm'}\widetilde{\Delta}(n,m)
\Bigg]\\
\end{split}
\label{eq:forP1}
\end{equation}
\begin{equation}
\begin{split}
\frac{\mathrm{d}}{\mathrm{d}t}P^{(2)}_{nmn'm'}=&
a(n+1)P^{(2)}_{n+1mn'-1m'}+b(n'+1)P^{(2)}_{n-1mn'+1m'}+\\
&A(m+1)P^{(2)}_{nm+1n'm'-1}+B(m'+1)P^{(2)}_{nm-1n'm'+1}-
\\
&anP^{(2)}_{nmn'm'}\widetilde{\delta}(n',m')-bn'P^{(2)}_{nmn'm'}\widetilde{\delta}(n,m)-\\
&AmP^{(2)}_{nmn'm'}\widetilde{\Delta}(n',m')-Bm'P^{(2)}_{nmn'm'}\widetilde{\Delta}(n,m)+
\\
\sum_{n''m''}\Bigg[
&an''P^{(3)}_{n''m''n-1mn'm'}+
bn''P^{(3)}_{nmn'-1m'n''m''}+\\
&a(n'+1)P^{(3)}_{nmn'+1m'n''m''}\widetilde{\delta}(n'',m'')+
b(n+1)P^{(3)}_{n''m''n+1mn'm'}\widetilde{\delta}(n'',m'')-\\
&an'P^{(3)}_{nmn'm'n''m''}\widetilde{\delta}(n'',m'')-
bnP^{(3)}_{n''m''nmn'm'}\widetilde{\delta}(n'',m'')-\\
&an''P^{(3)}_{n''m''nmn'm'}\widetilde{\delta}(n,m)-
bn''P^{(3)}_{nmn'm'n''m''}\widetilde{\delta}(n',m')\\
&Am''P^{(3)}_{n''m''nm-1n'm'}+
Bm''P^{(3)}_{nmn'm'-1n''m''}+\\
&A(m'+1)P^{(3)}_{nmn'm'+1n''m''}\widetilde{\Delta}(n'',m'')+
B(m+1)P^{(3)}_{n''m''nm+1n'm'}\widetilde{\Delta}(n'',m'')-\\
&Am'P^{(3)}_{nmn'm'n''m''}\widetilde{\Delta}(n'',m'')-
BmP^{(3)}_{n''m''nmn'm'}\widetilde{\Delta}(n'',m'')-\\
&Am''P^{(3)}_{n''m''nmn'm'}\widetilde{\Delta}(n,m)-
Bm''P^{(3)}_{nmn'm'n''m''}\widetilde{\Delta}(n',m')%
\Bigg]\;.
\end{split}
\label{eq:forP2}
\end{equation}
\end{widetext}
The factors $\widetilde{\delta}(n,m)$ and $\widetilde{\Delta}(n,m)$
ensure that the constraint (\ref{eq:stericbigsmall}) is satisfied
after addition of one small and big particle, respectively.
 {For general $k$
we have
\begin{equation}
\begin{split}
&\widetilde{\delta}(n,m)=\prod_{\overline{m}=0}^{\tilde{m}}
(1-\delta_{n(k-2\overline{m})}\delta_{m\overline{m}})\\
&\widetilde{\Delta}(n,m)=\prod_{\overline{m}=0}^{\tilde{m}}
(1-\delta_{n(k-2\overline{m})}\delta_{m\overline{m}})\times\\
&\quad\quad\quad\quad\quad\quad\quad(1-\delta_{n(k-1-2\overline{m})}\delta_{m\overline{m}})\\
\end{split}
\end{equation}
where $\tilde{m}=(k-1)/2$ for odd $k$ and $\tilde{m}=k/2$ for even $k$.
Note that in general case of arbitrary $k$ and arbitrary
(especially incommensurate) size parameters $c_1$ and $c_2$ the formulas
are more complicated and must be established on case-by-case basis. }

We are interested only in the stationary
state, so on the left-hand side of equations (\ref{eq:forP1}) and
(\ref{eq:forP2}), there are all zeros.
The mean-field approximation provides closure of the equations on
the single-site level, by setting
\begin{equation}
P^{(2)}_{nmn'm'}\simeq
P^{(2)}_{MFnmn'm'}=P^{(1)}_{nm}P^{(1)}_{n'm'}\;.
\label{eq:closure-mf}
\end{equation}
This yields a set of quadratic equations for one-site probabilities.
We already investigated the mean-field approximation in generalized
ASEP model in our previous work \cite{hum_kot_net_sla_20}. Here we aim at improving
the results using closure on two-site level, using a variant of the
Kirkwood approximation
\cite{kirkwood_35,kir_lew_ald_52,matsuda_00,sch_sch_nag_ito_95,mob_red_03,sla_szn_prz_08}.

 { Originally, it was developed for calculation
of properties of liquid mixtures \cite{kirkwood_35} and dense fluids
in general \cite{kir_lew_ald_52}. Essentially, it consists in
decoupling the three-point correlation function to a product composed
of two-site and one-site correlation functions.
It is worth noting
that the use of Kirkwood approximation in three-dimensional fluid
systems, like in  \cite{kirkwood_35},  and in one-dimensional discrete systems,
like in \cite{sch_sch_nag_ito_95,mob_red_03,sla_szn_prz_08}, differs
in important topological feature. Indeed, in 3D, all three points in
three-point correlation function are equivalent, and so the Kirkwood
decoupling formula must be symmetric with respect to permutation of
the triple. On the other hand, in 1D, one of the points lies in the
middle and two on the distant ends, and therefore the formula needs to
be symmetric just with respect to exchange of the two endpoints. Even
more, requirement of full permutation symmetry would ignore geometric
structure of the problem and thus must be avoided.
}

 {The formula for Kirkwood closure used by us is}
\begin{equation}
\begin{split}
&P^{(3)}_{nmn'm'n''m''}\simeq\\
&\;\;\;\;P^{(3)}_{Knmn'm'n''m''}=\frac{P^{(2)}_{nmn'm'}P^{(2)}_{n'm'n''m''}}{P^{(1)}_{n'm'}}\;.
\end{split}
\label{eq:closure-kirk}
\end{equation}
 {In fact, it is strictly equivalent to two-site cluster mean-field
approximations used in Refs.
\cite{sch_sch_93,pin_gov_13,hao_che_sun_liu_wu_16,cho_san_sch_00,mid_kol_gup_18}
but we shall keep the terminology naming it Kirkwood approximation in
our work.}

This leads to the set of algebraic equations for two-site
probabilities.
Note that the formula (\ref{eq:closure-kirk}) for Kirkwood approximation
is strictly equivalent to two-site cluster mean-field
approximations used in Refs.
\cite{sch_sch_93,pin_gov_13,hao_che_sun_liu_wu_16,cho_san_sch_00,mid_kol_gup_18}.

Moreover, the probabilities must satisfy a number of strict
constraints, which must not be violated in making the approximations
(\ref{eq:closure-mf}) or (\ref{eq:closure-kirk}). So, in the
mean-field approximation, we must impose the constraints
\begin{equation}
\begin{split}
&\sum_{mn}P^{(1)}_{nm}=1\\
&\sum_{mn}nP^{(1)}_{nm}=\rho_s\\
&\sum_{mn}mP^{(1)}_{nm}=\rho_b
\end{split}
\label{eq:mf-constraints}
\end{equation}
which effectively lower the number of quadratic equations to solve.
Similarly, in the Kirkwood approximations we impose the constraints
\begin{equation}
\begin{split}
&\sum_{mnm'n'}P^{(2)}_{nmn'm'}=1\\
&\sum_{mnm'n'}nP^{(2)}_{nmn'm'}=\sum_{mnm'n'}n'P^{(2)}_{nmn'm'}=\rho_s\\
&\sum_{mnm'n'}mP^{(2)}_{nmn'm'}=\sum_{mnm'n'}m'P^{(2)}_{nmn'm'}=\rho_b\\
&\sum_{m'n'}P^{(2)}_{nmn'm'}=\sum_{m'n'}P^{(2)}_{n'm'nm}\quad\quad\forall\; n,m\;.
\end{split}
\label{eq:kirk-constraints}
\end{equation}
Note, however, that the constraints (\ref{eq:kirk-constraints}) are
not all linearly independent. In Kirkwood approximation, there is
another set of constraints the two-site probabilities must satisfy,
namely the set of equations (\ref{eq:forP1}), which in stationary
state become just a set of linear dependencies between the two-site
probabilities. Again, not all of them are linearly independent, but
the relations (\ref{eq:kirk-constraints}) together with
(\ref{eq:forP1}) again effectively lower the number of algebraic
equations for two-site probabilities we must solve in the Kirkwood
approximation.

\begin{figure}[t]
\includegraphics[scale=0.85]{%
\slaninafigdir/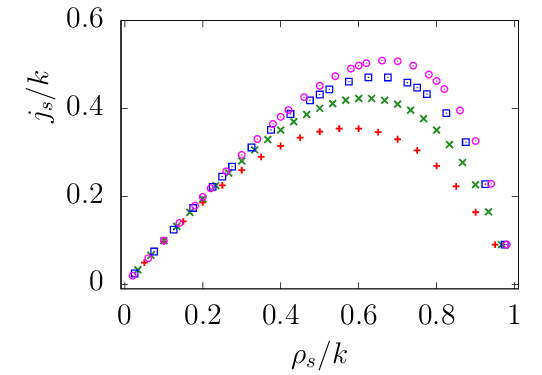}
\caption{Dependence of rescaled current on rescaled average density
for $a=1.5$, $b=0.5$, $\rho_b=0$ and for cell
capacities $k=2$ ($+$),$k=3$ ($\times$),$k=4$ ($\boxdot$),$k=5$ ($\odot$).}
\label{js-k2k3k4k5-rhob0-a1p5}
\end{figure}
\begin{figure}[t]
\includegraphics[scale=0.85]{%
\slaninafigdir/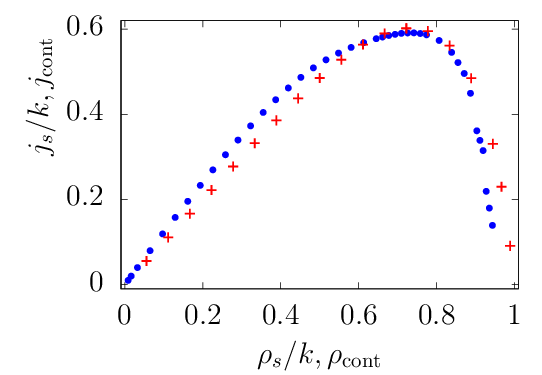}
\caption{ {Dependence of the current $j_\mathrm{cont}$
of particles   in the continuous model  (\ref{eq:stochcontinuous})
in a   two-dimensional channel of   width $w=3.1$ and length
$l=40$, on the density $\rho_\mathrm{cont}=N/w/l$   of particles
(symbols $\bullet$). The maximum number of particles   used was
$N=117$. As a comparison, the dependence of current $j_s$ on the
particle density $\rho_s$ in the one-dimensional generalized ASEP
model with   one type of particles, with $k=9$ (symbols $+$).}}
\label{js-driv-h}
\end{figure}

Let us show now how this general scheme works in the easiest case
$k=2$ with just small particles, i.e. with $\rho_b=0$. For simplicity
of notation, we omit the $m$ indices pertaining to big particles
everywhere. It is easy to see that the three equations for one-site
probabilities (\ref{eq:forP1}) are all identical, so there is only one
equation, namely
\begin{equation}
0=(a+b)P^{(2)}_{11}-2aP^{(2)}_{20}-2bP^{(2)}_{02}\;.
\label{eq:oneremaining}
\end{equation}
This is complemented by two constraints
\begin{equation}
\begin{split}
&P^{(1)}_{0} +P^{(1)}_{1} +P^{(1)}_{2}=1\\
&P^{(1)}_{1} +2P^{(1)}_{2}=\rho_s\;.
\label{eq:constraintsP1k2}
\end{split}
\end{equation}
In mean-filed approximation, this leads to a straightforward result
for, say, $P^{(1)}_1$, namely
\begin{equation}
P^{(1)}_1=-1+\sqrt{1+2\rho_s-\rho_s^2}
\end{equation}
and the remaining probabilities are computed using
(\ref{eq:constraintsP1k2}).

Up to now, the calculation was
trivial. The Kirkwood approximation is more involved. The set
(\ref{eq:forP2}) contains $9$ equations, but it can be easily seen
that only $7$ are linearly independent. The remaining two equations
correspond to first two constraints in (\ref{eq:kirk-constraints})
which fix the total probability to $1$ and the average density to
$\rho_s$. It can be also seen that summing the first, second and third
triples of the nine equations (\ref{eq:forP2}), we obtain exactly the
three equations for $P^{(1)}_n$, which are, as we already know, all
identical and impose the single constraint (\ref{eq:oneremaining}) on
the two-site probabilities. Altogether we find that the constraints
(\ref{eq:kirk-constraints}) together with the constraint
(\ref{eq:oneremaining}) form $5$ linearly independent relations for
$P^{(2)}_{nn'}$. But three of them are already contained in linear
combinations of equations (\ref{eq:forP2}). Therefore, we arrive at
two independent constraints (the first two in
(\ref{eq:kirk-constraints}))
and seven independent equations out of the set (\ref{eq:forP2}). This
makes consistently nine equations for nine unknown probabilities
$P^{(2)}_{nn'}$. These equations become closed and algebraic, as soon
as we apply the Kirkwood approximation (\ref{eq:closure-kirk}). These
equations must be solved numerically.
 {We used an iteration procedure
which emulates the time evolution of the two-site correlation functions,
written in the equations (\ref{eq:forP2}). The initial condition was
taken as the solution within mean-field approximation. Then, in each
step we take the two-site probabilities we have, and using the
Kirkwood closure, insert them into the right-hand side of
(\ref{eq:forP2}). The left-hand side of (\ref{eq:forP2})
provides corrections to the two-site probabilities for the next step.
This procedure is iterated until stationarity.
From a practical point of view,
we found that such iteration is } reliable and numerically
stable, if we proceed by iterations of all the algebraic equations
(\ref{eq:forP2}) and in each step we correct the rounding errors by
imposing all linear constraints (\ref{eq:kirk-constraints}) and
(\ref{eq:oneremaining}) explicitly. This way the system of equations is formally
overdetermined, which helps to cure numerical instabilities.
We used the same technique in all calculations shown below.

\begin{figure}[t]
\includegraphics[scale=0.85]{%
\slaninafigdir/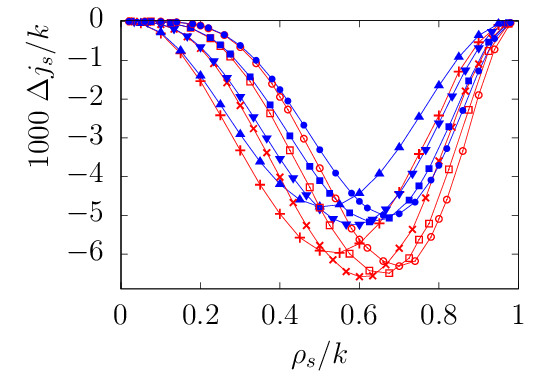}
\caption{Dependence of the difference in rescaled current on the
rescaled average density, for $a=1.5$, $b=0.5$.
The differences between numerical simulation data and mean-field
results are denoted by symbols $+$ ($k=2$), $\times$ ($k=3$), $\boxdot$ ($k=4$),
and $\odot$  ($k=5$).
The differences between Kirkwood approximation results and mean-field
results are denoted by symbols
$\blacktriangle$ ($k=2$), $\blacktriangledown$ ($k=3$), $\blacksquare$ ($k=4$),
and $\bullet$  ($k=5$).
}
\label{js-differences-k2k3k4k5-rhob0-a1p5}
\end{figure}

\section{Nearest-neighbor correlations and their effect on  particle currents}

The current of particles ($j_\alpha$ for particle type $\alpha$) is
uniquely determined by two-site
probabilities
\begin{equation}
\begin{split}
j_s=\sum_{nmn'm'}\Big(&naP^{(2)}_{nmn'm'}\widetilde{\delta}(n',m')-\\
&n'bP^{(2)}_{nmn'm'}\widetilde{\delta}(n,m)\Big)\\
j_b=\sum_{nmn'm'}\Big(&mAP^{(2)}_{nmn'm'}\widetilde{\Delta}(n',m')-\\
&m'BP^{(2)}_{nmn'm'}\widetilde{\Delta}(n,m)\Big)\;.
\end{split}
\label{eq:currents}
\end{equation}
 {Therefore, the currents are the first and immediate
indicators of the correlations present in the system. More
specifically, the correlations are manifested in the deviation of
the actual current from the mean-field
predictions (\ref{eq:closure-mf})  and various
approximations, in our case the Kirkwood one, can be tested using
this deviation.}

Let us first show the density dependence of current in one-component
system (small particles only) for different cell capacities $k$
(Fig. \ref{js-k2k3k4k5-rhob0-a1p5}), as obtained by numerical Monte
Carlo simulations.
 {(The details of the simulation procedure are as
follows. The initial condition is completely random configuration of
particles, the typical length of the one-dimensional system is
$L=10000$, each run typically consists of $10^5$ thermalization steps
followed by $10^5$ steps in which current and correlations are
measured. Each step consists of $L$ elementary one-particle
updates. The results are averaged over typically $100$ independent
runs and the statistical error is smaller than symbol size in all
figures shown.) }
From the data for $k\le 5$ we can see that the
position of the maximum on the axis of rescaled density $\rho_s/k$
increases with $k$ and the whole curve approaches the
independent-particle limit for $k\to\infty$. This stresses the
necessity of rational choice of the cell size in the local mixing
approximation. Indeed, the reasonable procedure would be to compare
the current-density chart of the original continuous system (whatever
its source may be, empirical, numeric simulations of the model
(\ref{eq:stochcontinuous}), (\ref{eq:stericcontinuous}), or else) with
numerical simulations like those in Fig.
\ref{js-k2k3k4k5-rhob0-a1p5} and choose $k$ and factors $c_\alpha$
according to best coincidence.

 {
To show an example, we
performed a small-scale two-dimensional simulation of the process
(\ref{eq:stochcontinuous}) with constraints
(\ref{eq:stericcontinuous}) in a straight channel, with just one type
of particles. The particles were
hard disks  with unit diameter. From the
simulations, we extracted the current-density diagram. In this case,
the density is defined as $\rho_\mathrm{cont}= N/w/l$ for $N$
particles in a channel of width $w$ and length $l$. In our case, the
channel width was $w=3.1$,
i.e. three particles entered side-by-side into the channel.
We found, that
it best fits with the current-density diagram of the generalized ASEP
process with $k=9$. The match is shown in Fig. \ref{js-driv-h}.
So, in this specific case we can conclude that the proper choice is
$k=9$. Of course, more complex geometries and mixtures of particles of
different sizes may require more thorough investigation. It also can
be expected that in complicated geometries the cell size and capacity
may not be uniform. However, here we do not intend to go to such
details and investigate just generic properties of the model with
fixed $k$. We shall also mostly concentrate on the simplest
non-trivial case $k=3$.}

To see the effect of correlations, we take the values of the current
calculated in mean-filed approximation (\ref{eq:closure-mf}) as a
reference and subtract it from the value obtained by numerical
simulations on one side and by Kirkwood approximation
(\ref{eq:closure-kirk}) on the other side. This way we can assess, to
what extent the Kirkwood approximation grasps the correlations
actually present in the system.

We can compare the differences in current for one-component system and
cell capacities from $k=2$
to $k=5$ in Fig. \ref{js-differences-k2k3k4k5-rhob0-a1p5}. First of
all, we note that the absolute value of the difference is small. It
amounts just a few per cent of the value of the current. Next, the
size of the difference, when measured in rescaled current $\rho_s/k$,
is only weakly dependent on $k$. It seems that it is largest for
$k=3$ and slowly decreases for increasing $k$,  but the decrease is
indeed very slow. Interestingly, the trend is identical both in
numerical simulations and in Kirkwood approximation. The fact that
correlations gradually vanish if we approach either zero density or
maximum density is obvious: these are regimes of independent particles
and independent holes, respectively. Note, however, that there is no
particle-hole symmetry in the model, so the process of nearly independent holes for
$\rho_s/k\to 1$ cannot be easily mapped on a process of nearly
independent particles.

\begin{figure}[t]
\includegraphics[scale=0.85]{%
\slaninafigdir/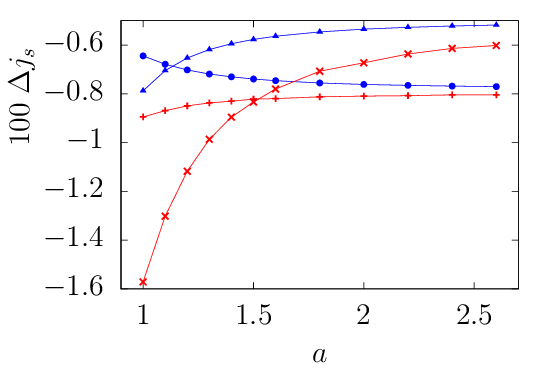}
\caption{Dependence of the difference in current on the hopping rate
$a$, for $k=3$, with fixed $a-b=1$.
The differences between numerical simulation data and mean-field
results are denoted by symbols $+$ ($\rho_s=1$) and $\times$ ($\rho_s=2.5$).
The differences between Kirkwood approximation results and mean-field
results are denoted by symbols  $\bullet$  ($\rho_s=1$)
and $\blacktriangle$ ($\rho_s=2.5$).}
\label{js-differences-vs-a-k3-rohs1rhos2p5-rhob0}
\end{figure}

Comparing the values for numerical simulations with those from
Kirkwood approximation, we can see that Kirkwood approximation
explains most of the correlation effect, but certainly not all. We can
see how this comparison looks when we change the hopping rates $a$,
$b$.
 {
In Fig. \ref{js-differences-vs-a-k3-rohs1rhos2p5-rhob0},
we show the dependence on $a$, with drift $a-b$ kept
constant, for two fixed densities.  We can clearly see that for larger
$a$ (and therefore larger
diffusion coefficient $D=(a+b)/2$) the agreement between
Kirkwood and numerical simulations is better. This observation is valid
at all densities.  We can also see that
this is not due to decrease in correlations themselves. The amount of
correlations, as measured by the difference from mean-filed data,
approaches a constant definitely distinct from zero, when $a$ increases.
}


%

\begin{figure}[t]
\includegraphics[scale=0.85]{%
\slaninafigdir/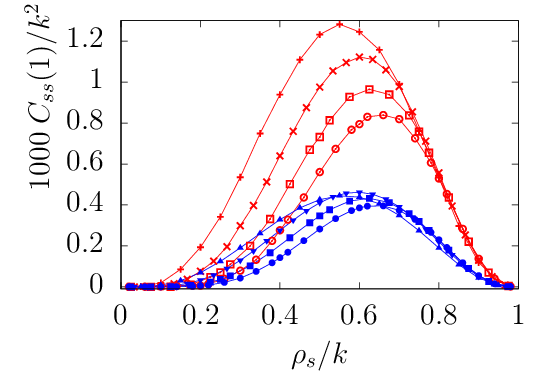}
\caption{Dependence of rescaled nearest-neighbor density-density
correlation on rescaled average density. The hopping rates are $a=1.5$,
$b=0.5$. The numerical simulation results are denoted by symbols
$+$ ($k=2$), $\times$ ($k=3$), $\boxdot$ ($k=4$),
and $\odot$  ($k=5$).
The corresponding Kirkwood approximation results are
denoted by symbols
$\blacktriangle$ ($k=2$), $\blacktriangledown$ ($k=3$), $\blacksquare$ ($k=4$),
and $\bullet$  ($k=5$).
}
\label{cor-nn-k2k3k4k5-rhob0-a1p5}
\end{figure}

Particle current is an indirect indicator of correlations, but we can
see the correlations directly in density-density correlation function
\begin{equation}
\begin{split}
C_{\alpha\beta}(x) = \langle &n_\alpha(x')n_\beta(x'+x)\rangle-\\
&\langle
n_\alpha(x')\rangle\langle n_\beta(x'+x)\rangle\;.
\end{split}
\end{equation}
Here the indices $\alpha$, $\beta$ are either $s$ or $b$ denoting
small and big particles, respectively. The nearest neighbor
correlations ($x=1$) in the system composed of only small particles
are shown in Fig. \ref{cor-nn-k2k3k4k5-rhob0-a1p5} for cell capacities
from $k=2$ to $k=5$. The density is rescaled by $k$ in order to obtain
comparable results. We can see that correlations measured by
density-density correlation function decreases clearly when $k$
increases, as we should expect, because for $k\to\infty $ particles
are independent. This contrasts with the finding for particle
current. Recall that the decrease of correlations, as measured by
current difference,  with $k$, was rather slow.

\begin{figure}[t]
\includegraphics[scale=0.85]{%
\slaninafigdir/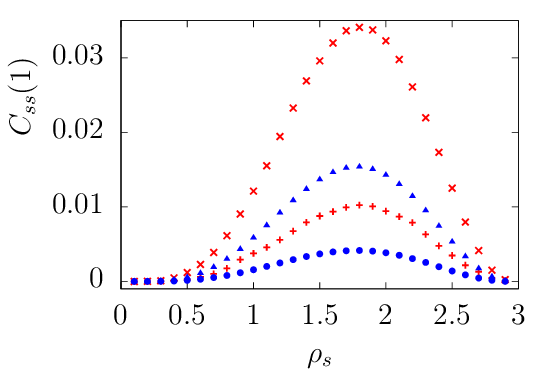}
\caption{Dependence of nearest-neighbor density-density
correlation on average density, for $k=3$.  The numerical simulation results
are denoted by symbols
$\times$  ($a=1$, $b=0$) and
$+$  ($a=1.5$, $b=0.5$).
The corresponding Kirkwood approximation results are
denoted by symbols
$\blacktriangle$  ($a=1$, $b=0$) and
$\bullet$ ($a=1.5$, $b=0.5$).
}
\label{cor-nn-k3-rhob0-a1p5a1}
\end{figure}

Similarly, in Fig. \ref{cor-nn-k3-rhob0-a1p5a1} we can compare the
Kirkwood and simulation results for different values of hopping
rates.
 {We can see that the correlations are weaker when the diffusion
coefficient $D=(a+b)/2$ is larger, at fixed driving $a-b$. This is reproduced
within the Kirkwood approximation. On the other hand, we observed that
the Kirkwood approximation underestimates the density-density
correlations by a factor about two, independently of $a$. Currently,
we do not have any easy explanation of this systematic deviation.}

\begin{figure}[t]
\includegraphics[scale=0.85]{%
\slaninafigdir/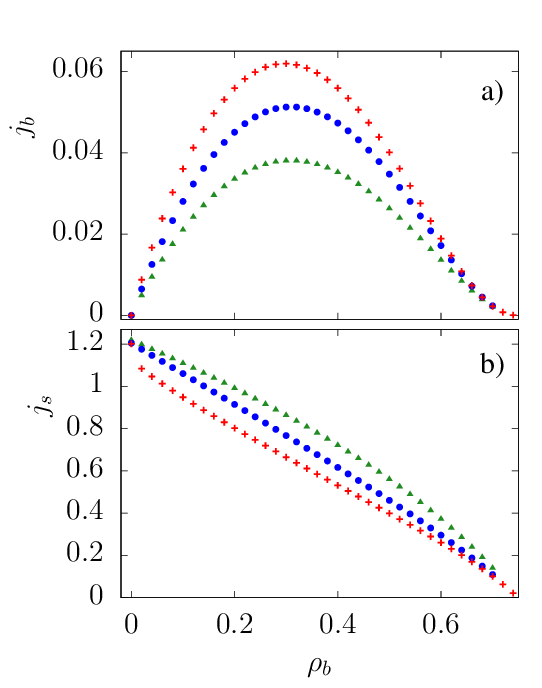}
\caption{Dependence of current of big (panel a)) and small (panel b))
particles on the
average density of big particles, in two-species system, for $k=3$
$\rho_s=1.5$ and rates $a=1.5$, $b=0.5$, $A=0.75$, $B=0.25$. The
results of numerical simulations are denoted by symbol $+$, results
of mean-field approximation by $\blacktriangle$ and results of
Kirkwood approximation by $\bullet$.
}
\label{jsjb-k3-rhos1p5-a1p5}
\end{figure}

Finally, we investigated the correlations in mixed system, with small
concentration of big particles. Let us first look at how the
correlations manifest in the current of small and big particles. We
can compare the simulations, mean-field and Kirkwood results in
Fig. \ref{jsjb-k3-rhos1p5-a1p5}. Clearly, Kirkwood approximation is
better than mean-field approximation by about 50 per cent. However,
there is a feature which neither mean-field nor Kirkwood
approximations grasp correctly, namely the behavior of the current of
small particles at very small concentration of big particles. Indeed,
the simulation data show that even addition of a single big particle
in a system of size $L=10000$ sites causes perceptible drop in the
current of small particles $j_s$ from its $\rho_b=0$ value (we
investigated this effect already in our previous work
\cite{hum_kot_net_sla_20}). On the
other hand, both mean-field and
Kirkwood approximation predict that $j_s$ approaches its $\rho_b=0$
limit gradually, without any sudden drop. This implies that admixture
of big particles among small ones induces long-range correlations
which
 {cannot be reproduced within Kirkwood approximation.
We found that this phenomenon is related to the fact that larger particles
have smaller hopping rates, corresponding to the properties of
Brownian particles, namely that diffusion coefficient is inversely
proportional to particle diameter.
Indeed, to show the difference, we also investigated the (unphysical) case
when larger particles have larger hopping rates, rather than
smaller. In this case, the current-density diagrams analogous to
Fig. \ref{jsjb-k3-rhos1p5-a1p5} look qualitatively different. Most
importantly, small
admixture of big particles among small ones
makes very little difference. Moreover, we found that the Kirkwood
approximation fits
the simulation data very well for all densities. This suggests that
the long-range correlations, unexplained by the Kirkwood
approximation, are caused by slower movement of larger particles.
}

\begin{figure*}[t]
\includegraphics[scale=0.85]{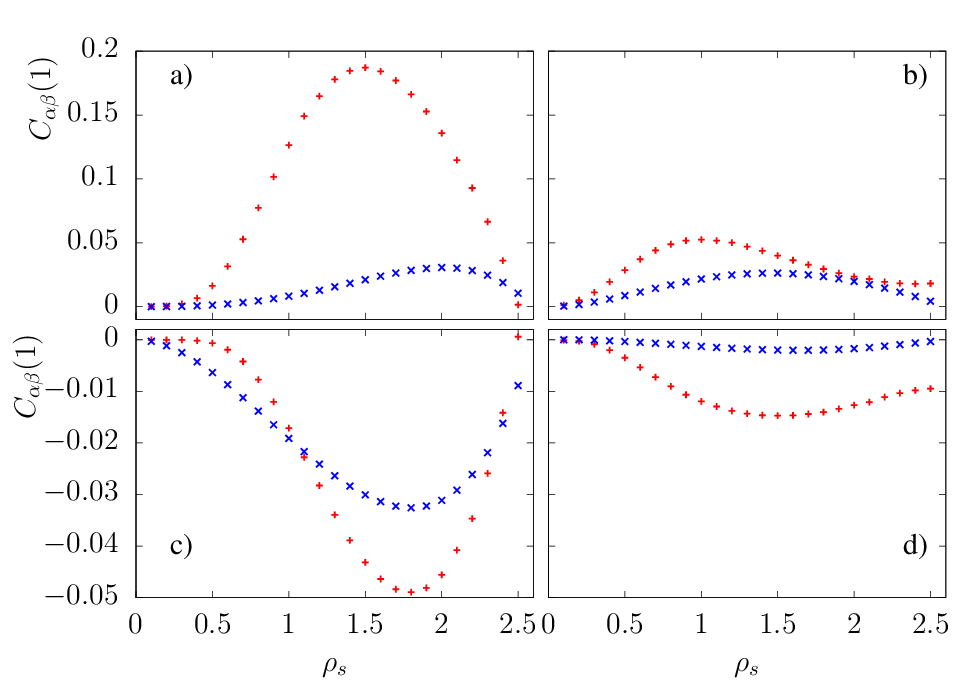}
\caption{Dependence of nearest-neighbor correlation functions on the
density of small particles. The density of big particles is
$\rho_b=0.2$. The parameters of the model are
$k=3$, $a=1.5$, $b=0.5$, $A=0.75$, and $B=0.25$. The symbols
distinguish the numerical simulations ($+$) and the Kirkwood
approximation ($\times$). The four panels correspond to four
combinations of the particle types; in panel a) $\alpha\beta=ss$,
in  panel b)  $\alpha\beta=sb$,
in panel c)  $\alpha\beta=bs$, and in   panel d)  $\alpha\beta=bb$.
}
\label{cor-nn-k3-rhob0p2-a1p5-b0p5}
\end{figure*}

We can also look at the nearest-neighbor correlation functions
directly. In
Fig. \ref{cor-nn-k3-rhob0p2-a1p5-b0p5}
we can compare the results of simulations with Kirkwood
approximation. For the small-small correlation function, we observe
that Kirkwood approximation largely underestimates the actual
value. This can be compared with the results for one-component system
shown in Fig. \ref{cor-nn-k3-rhob0-a1p5a1} where the agreement was
better. This implies that the presence of big particles induces
enhanced correlation among small particles. The picture becomes clear
from the mixed small-big and big-small correlations, as can be seen in
Fig.  \ref{cor-nn-k3-rhob0p2-a1p5-b0p5}. The correlation function
$C_{sb}(1)$ is positive, while $C_{bs}(1)$ is negative. This can be
easily understood, as the big particle blocks the movement of small
particles behind it, increasing locally their density, while in front
of the big particle the density of small particles is diminished. We
can also see that the correlations among big particles, due to the presence
of small particles, is negative. This can be also interpreted as
consequence of accumulation of small particles behind a big
particle. Indeed, the pure system of big particles is in fact
identical to original ASEP, therefore correlations between big
particles are absent. If we include some small particles, they
accumulate behind big ones and impede other big particles to come
closer. Small particles act as spacers which keep big particles
apart. Hence the negative autocorrelation of big particles.
Comparing the Kirkwood and simulation data, we again observe the
Kirkwood approximation largely underestimates the autocorrelation of
big particles, exactly as it does in the case of autocorrelation of
small particles. Moreover, there is significant qualitative difference
between Kirkwood and simulation data. When the density of small
particles approaches its maximum dictated by the density of big
particles, the correlation functions  $C_{sb}(1)$  and $C_{bb}(1)$
approach a non-zero limit, while Kirkwood approximation wrongly predicts
approach to zero. As a summary, we conclude that the case of mixture
of small and big particles is only very poorly described by the
Kirkwood approximation.

\section{Long-range correlations}

\subsection{Implications of Kirkwood approximation}

Beyond the next-neighbor correlations, we can use the result for
Kirkwood approximation in a chain-like form. If we define
\begin{equation}
\begin{split}
P^{(2)}_{nmn'm'}(x)=\frac{1}{L}\sum_{x'=1}^L
\mathrm{Prob}\{n_s(x')=n,n_b(x')=m,\\
n_s(x'+x)=n',n_b(x'+x)=m'\}
\end{split}
\end{equation}
we can express it in Kirkwood approximation as
\begin{equation}
\begin{split}
P^{(2)}_{nmn'm'}(x)\simeq
\sum_{n_1m_1}&\big[M^{x-1}\big]_{nmn_1m_1}P^{(2)}_{n_1m_1n'm'}
\end{split}
\end{equation}
where the matrix
\begin{equation}
M_{nmn'm'}=\frac{P^{(2)}_{nmn'm'}}{P^{(1)}_{n'm'}}
\end{equation}
is stochastic, i.e. it has an eigenvalue equal $1$. The largest
eigenvalue $\lambda$ lower than $1$ governs the large-distance decay
of the density-density correlation function
\begin{equation}
C_{\alpha\beta}(x) \sim \lambda^{x}\;,\quad x\to\infty \;.
\label{eq:exonentialdecay}
\end{equation}
We can see that the structure of the Kirkwood approximation
necessarily implies exponential decay of correlations. Depending on
the dimension of the matrix, which is given by the number of
allowed one-site configurations, there are several sub-leading
exponential contributions, which may be responsible for alternating
sign of the correlation function, but do not influence the general
exponential character. This observation also explains why the decay of
correlation was always exponential in those generalizations of ASEP,
in which the two-site cluster approximation (equivalent to the
Kirkwood approximation as used in our work) is in fact exact
\cite{gab_kra_red_10,bas_moh_09}.
In this Section, we shall show that in our generalized ASEP model the
actual decay of correlation functions is fundamentally different
from (\ref{eq:exonentialdecay}).

\subsection{One-component system}

\begin{figure}[t]
\includegraphics[scale=0.85]{%
\slaninafigdir/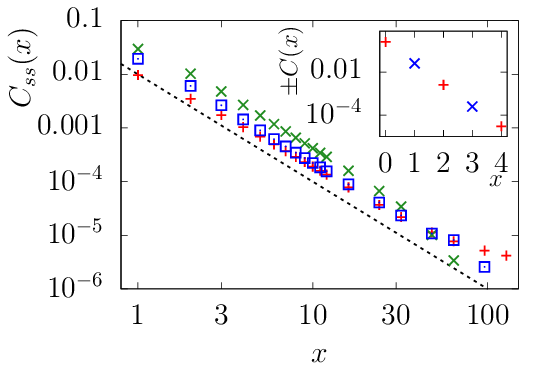}
\caption{Long-range density-density correlations for $\rho_b=0$, $\rho_s=1.5$
$a=1$, $b=0$. Different symbols correspond to
$k=2$,($ +$),
$k=3$, ($ \times$),
$k=4$, ($\boxdot $).
The dotted line is the power $\propto x^{-2}$. In the inset,
density-density correlation function for the facilitated ASEP model, with
parameter $f=1.5$. The symbols distinguish the sign in front of the quantity
$C$:  symbol $+$ corresponds to the plus sign (i.e. correlation
function is positive)   and symbol $\times$ corresponds to the minus
sign (i.e. correlation function is negative).
}
\label{cor-long-k2k3k4-rhos1p5-rhob0}
\end{figure}

We simulated the system composed of small particles only, for $k=2$,
$k=3$, and $k=4$.  The results for density $\rho_s=1.5$ are shown in
Fig. \ref{cor-long-k2k3k4-rhos1p5-rhob0}. We can see that the
density-density correlation function is nearly independent of the
cell capacity $k$ and decays as a power law with exponent close to
$2$. For comparison, we performed also simulations of the facilitated
ASEP model \cite{gab_kra_red_10,bas_moh_09}. The density-density
correlation function is shown in the inset of Fig.
\ref{cor-long-k2k3k4-rhos1p5-rhob0} and we can see that the decay is
exponential, with regularly alternating sign. This is in stark
contrast with the results of our model and shows that our
generalization of ASEP belongs to different universality class than
the set of models investigated in e.g. \cite{hao_jia_hu_jia_wan_16}
where the authors conclude that exponential decay of correlations is
universal
property.

\begin{figure}[t]
\includegraphics[scale=0.85]{%
\slaninafigdir/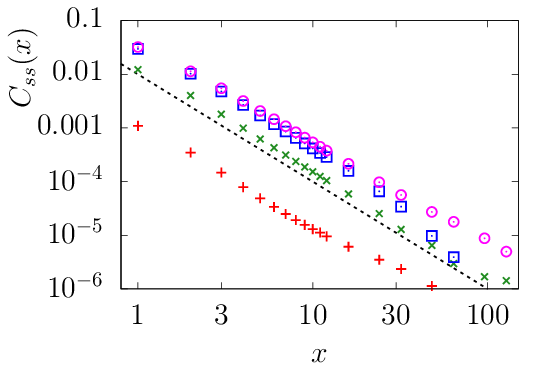}
\caption{Long-range density-density correlations for $k=3$, $\rho_b=0$,
$a=1$, $b=0$. Different symbols correspond to
$\rho_s=0.5$,($ +$),
$\rho_s=1$, ($ \times$),
$\rho_s=1.5$, ($\boxdot $),
$\rho_s=2$, ($\odot $).
The dotted line is the power $\propto x^{-2}$.}
\label{cor-long-k3-rhob0}
\end{figure}

We also checked the dependence of correlation function on particle
density. The results of simulations is shown in
Fig. \ref{cor-long-k3-rhob0}. We can see that the amplitude of
correlations does depend on the density, but the algebraic character
of decay is unchanged. For all densities studied we observe power-law
decay with the same exponent close to $2$.
 {We also checked that this behavior is independent of
the choice of hopping rates $a$ and $b$.}
Therefore, we can
conjecture that the one-component generalized ASEP is universally
characterized by power-law decay of correlations wit exponent equal to
$2$. At present we do not have an analytic argument which would imply
such universal value of the exponent.


\begin{figure*}[t]
\includegraphics[scale=0.85]{%
\slaninafigdir/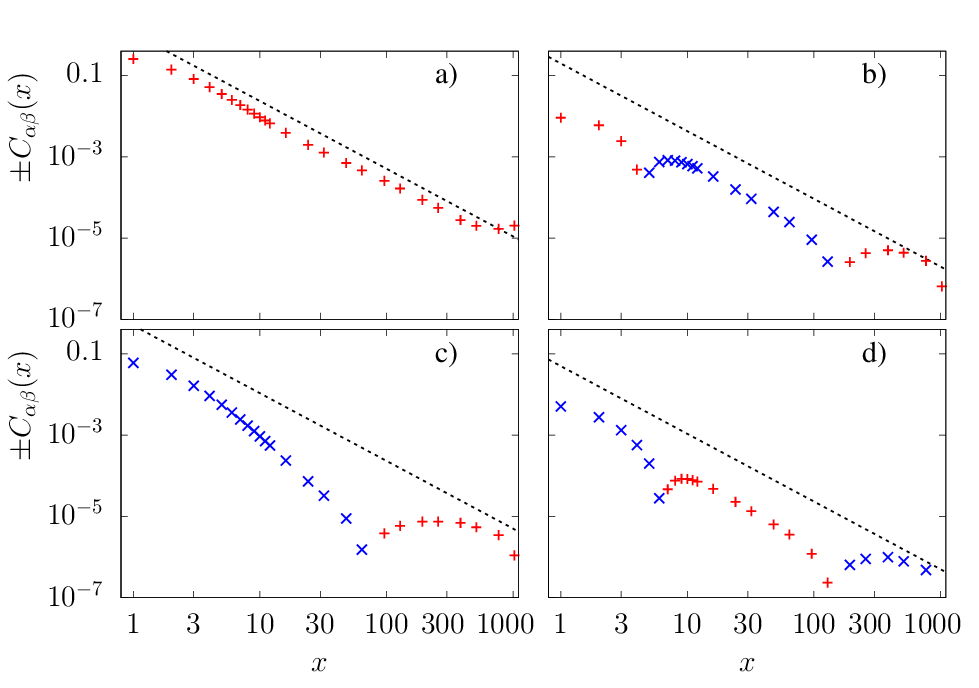}
\caption{ Long-range density-density correlations in two-species
system, $k=3$, with
densities $\rho_s=2$, $\rho_b=0.1$, $a=1$, $A=0.5$, $b=B=0$. The
four panels correspond to
pairs $\alpha\beta\equiv ss$ (panel a)),
$\alpha\beta\equiv sb$ (panel b)),
$\alpha\beta\equiv bs$ (panel c)),
$\alpha\beta\equiv bb$ (panel d)).
The symbols distinguish positive and negative parts of the
correlation function, indicated by the prefix $\pm$. Symbols
$+$ correspond to $+$ sign, symbols $\times$ correspond to $-$
sign. The dotted line is the power-law decay $\propto x ^{-5/3}$.
}
\label{cor-long-k3-rhos2-rhob0p1}
\end{figure*}

\begin{figure}[t]
\includegraphics[scale=0.85]{%
\slaninafigdir/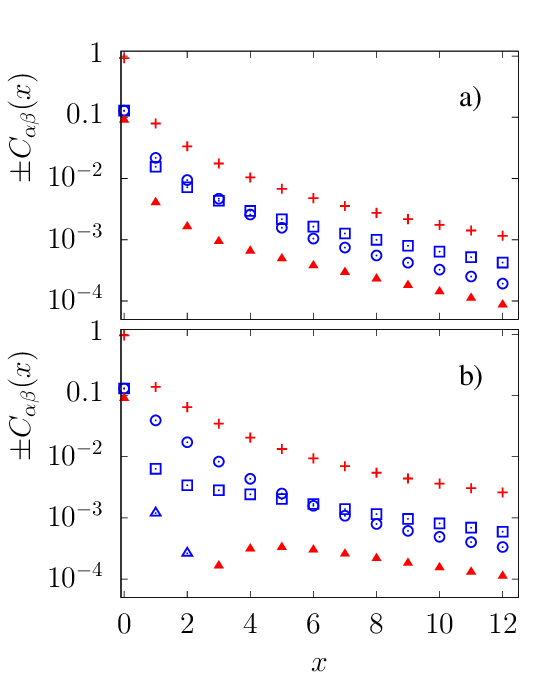}
\caption{  {Long-range density-density correlations in two-species
system, $k=3$, with
densities $\rho_s=2$, $\rho_b=0.1$, and rates $a=0.5$, $A=1$,
(panel a)), and  $a=A=1$ (panel b)). In both cases $b=B=0$.
The symbols correspond to
the following pairs and signs in front of the correlation function:
$\alpha\beta\equiv ss$, sign $+$ (symbol $+$),
$\alpha\beta\equiv sb$,  sign $-$ (symbol $\boxdot$),
$\alpha\beta\equiv bs$,  sign $-$ (symbol $\odot$),
$\alpha\beta\equiv bb$,  sign $+$ (symbol $\blacktriangle$),
$\alpha\beta\equiv bb$,  sign $-$ (symbol
$\bigtriangleup\!\!\!\cdot\;$).}
}
\label{cor-long-k3-rhos2-rhob0p1-as0p5-ab1}
\end{figure}
\begin{figure}[t]
\includegraphics[scale=0.85]{%
\slaninafigdir/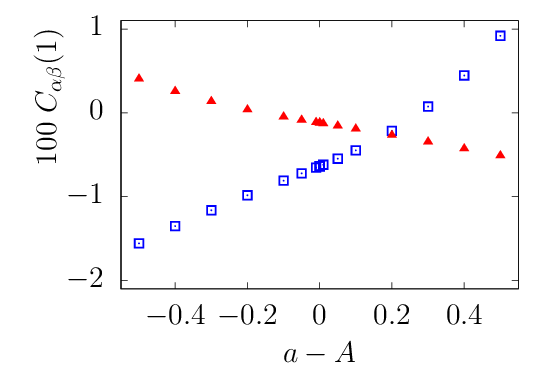}
\caption{ {Dependence of the nearest neighbor
density-density
correlations in a two-species system, $k=3$, with
densities $\rho_s=2$, $\rho_b=0.1$, on the difference of hopping
rates of the small
and big particles. The rates $b=B=0$ are fixed. For $a\le 1$ we fix
$A=1$, for $A<0$ we fix $a=1$. The correlation functions shown are
$\alpha\beta\equiv sb$ (symbol $\boxdot$), and
$\alpha\beta\equiv bb$ (symbol $\blacktriangle$).}
}
\label{cor-nn-vs-a-k3-rhos2-rhob0p1}
\end{figure}

\subsection{Two-component system}

Now we ask how the picture of long-range correlations changes when we
mix among small particles also a portion of big particles. A typical
result is shown in Fig. \ref{cor-long-k3-rhos2-rhob0p1}. We can see
that the correlation function $C_{ss}(x)$ of small particles    among
themselves decays again as a power law, but with somewhat smaller
exponent. We estimated the exponent to about $5/3$. Other
correlation functions, i.e. the big-big $C_{bb}(x)$ and mixed
correlations $C_{sb}(x)$ and $C_{bs}(x)$ do not decay
monotonously, but change sign. However, the sign does not
change regularly, as in the facilitated ASEP (see the inset in
Fig. \ref{cor-long-k2k3k4-rhos1p5-rhob0}), but there are protracted
intervals of constant sign. Also, the distances at which the
sign change takes place are not the same for all three
correlation functions  $C_{bb}(x)$ and mixed correlations $C_{sb}(x)$
and $C_{bs}(x)$. We found such type of behavior for all densities of
small and big particles studied, as soon as there is non-negligible
density of big particles. But in spite of not much transparent pattern
of sign changes, it seems that the envelope of the correlation
function again decays very slowly and the decay is compatible with
power law with the same exponent $5/3$ as in the decay of small-small
correlation function.

 {
With passive Brownian particles, both the diffusion coefficient and
mobility are inversely proportional to particle size. Hence, we
naturally assumed that the hopping rates of big particles are smaller
than the hopping rates of small particles. The results discussed so
far rely on this assumption.
For comparison, we also studied the (unphysical) case in which the
large particles have larger diffusion coefficient and larger
mobility than the smaller particles. We show the results in
Fig. \ref{cor-long-k3-rhos2-rhob0p1-as0p5-ab1}a. The picture is
rather different from the usual case, in which larger particles are
slower. Indeed, comparison with Fig. \ref{cor-long-k3-rhos2-rhob0p1}
reveals, that in absolute value the correlations are very similar in
the two cases, but the sign differ. As we have seen, slower big particles cause
accumulation of small particles behind and rarefaction in front of
a big particle, as demonstrated by positive small-big correlations
and negative big-small correlations. At the same time, big particles
effectively repel each other. On the other hand, when the big
particles are faster than the small
ones, this effect does not happen. Instead, the mixed correlations
of type small-big and big-small have both negative sign and the
correlation bib-big has positive sign. This means that here the
interaction between small and big particles leads to separation of
particle types. Effectively, the big and small repel each other,
while big effectively attract big and small effectively attract
small. This is a completely different regime of the
dynamics. }

 {We also looked at the question of how the regime of
slower big particles transits to the regime of slower small
particles. For example, we show in
Fig. \ref{cor-long-k3-rhos2-rhob0p1-as0p5-ab1}b  the
correlation function in the case of equal hopping rates of small and
big particles, $a=A$. We can see that the behavior is indeed a kind of
mixture. The small-big correlations are still negative as in the
case $a<A$, but bib-big correlations are negative at short
distances, as in the case $a>A$. Another insight provides
Fig. \ref{cor-nn-vs-a-k3-rhos2-rhob0p1}, where we show the
nearest-neighbor correlations as function of the difference of
the hopping rates of small and big particles. Important conclusion
is, that the dependence is smooth, so there is no sharp transition
between regimes. Another interesting observation is that the
correlation functions change sign at a general value of $a-A$,
rather than at $a=A$, as one could naively expect.}

 {However, we would like to stress that
it remains somewhat speculative, how to
practically realize
faster diffusion and driving with larger particles. One such
possibility would follow from the use of active particles
\cite{bec_dil_low_rei_vol_vol_16} but this goes beyond the scope of this work.}

\section{Conclusions}

We investigated correlation effects in a generalized ASEP model with
two types of particles (called small and big) and cell capacity $k$
larger than one. For
nearest-neighbor correlations we used the Kirkwood
approximation. Comparing this approximation with numerical simulations
we found fairly good agreement as long as there are only small
particles. In the mixed system with both small and big particles, we
observed more complicated picture. Even the presence of very small
concentration of big particles among the small ones changes the
behavior substantially. The small particles accumulate behind the big
ones, which leads to substantial decrease of current. This behavior is
not grasped by the Kirkwood approximation. Also the density-density
autocorrelation function of small particles is largely underestimated
by Kirkwood approximation, as soon as there is admixture of big
particles. However, Kirkwood approximation still gives qualitatively
correct results, notably it correctly predicts accumulation of small
particles behind the big ones and lowered density of small in front of
big particles. It also correctly predicts negative autocorrelation of
big particles in presence of small particles, but again,
quantitatively the Kirkwood approximation largely underestimates the
size of the correlation function.

We also thoroughly investigated long-range correlations in the
generalized ASEP model. We showed that if the Kirkwood approximation
was reasonably accurate, the density-density correlation function
would decay exponentially. This is correct in the facilitated ASEP
model, where Kirkwood approximation is in fact exact and the
correlation function indeed decays exponentially. However, in our
generalization of ASEP model, the correlations decay algebraically,
rather than exponentially. Interestingly, for one-component system
with small particles only, the exponent of the power-law decay seems
to be universal and equal to $2$,
 {independently of  the average
particle density, on the cell capacity
$k$ and on the hopping rates.}
This means that our generalization of ASEP model belongs to
different universality class than for example the facilitated ASEP or
the whole set of ASEP generalizations which can be mapped on
zero-range processes. We do not have any simple argument which would
explain the exponent $2$ observed in our model. It seems that a new
analytic approach is necessary here.

 {
Interestingly, a similar power-law decay of correlations with same
exponent has also been observed in the two-dimensional lattice gas
subject to a strong driving field in one direction, while in the other
direction the particles are ruled by the high-temperature Kawasaki
dynamics \cite{zha_wan_leb_val_88}. By a perturbative argument, the
correlations therein are
shown to be dominated by a quadrupole solution to the two-dimensional
Laplace equation, which explains the decay exponent equal two. It
remains an open question whether a theoretical explanation for the
similar correlation decay in both driven models can be given, or
whether it is mere coincidence.  }

In mixed case of small and big particles together, the situation is
modified. When we look at the autocorrelation of small particles, the
correlation function decays again algebraically, with somewhat smaller
exponent than for one-component system. The mixed small-big and
big-small correlation functions as well as the autocorrelation
function of big particles  decay non-monotonously and exhibit
change of sign. However, the envelope of the oscillating correlation
functions seem to decay again algebraically, although it is not easy
to deduce the exponent. In any case, however, the data show much
slower decay than exponential. The sign of mixed correlation functions
again confirms accumulation of small particles behind the large ones
and decrease of concentration in front of big particles. There is also
interesting asymmetry: the dense region behind the big particle seems
to be much shorter than the rare region in front of the big
particle. This can be interpreted in the following way. Behind a big
particles, there is a jam of small particles, which in turn leads to
accumulation of other big particles. This shortens the region filled
by small particles. In front of a big particle, no such mechanism is
at work, and larger interval with lower density of both small and big
particles is observed. This is also consistent with the behavior of
autocorrelation of big particles. At short distances, the
autocorrelation is negative, only at larger distances it changes sign
to positive. Note that
 {for $k\le 3$}
the one-component system of big particles only
is equivalent to original ASEP, so the correlations are totally
absent.
 {(For $k>3$ it is no more true).}
The non-trivial autocorrelation of big particles observed in
the simulations is entirely due to the presence of small
particles. And the fact that at short distance the autocorrelation is
negative implies that the small particles act as spacers, which keep
the big particles far to each other.

To sum up, we showed that the generalization of ASEP studied here
belongs to different universality class than generalizations studied
earlier. Furthermore, mixing big and small particles adds further complexity
which is only poorly described by usual Kirkwood approximation.

\begin{acknowledgments}
We wish to thank Y. Humenyuk and P. Kalinay for inspiring discussions.
\end{acknowledgments}
%
%
%
%
%
%

%
%
%
%
\end{document}